\documentclass{iau-JDSS}

\usepackage{graphicx}

\title[Similar phenomena at different scales] 
{Similar phenomena at different scales: Black Holes, the Sun, Gamma-ray Bursts,
Supernovae, Galaxies and Galaxy Clusters}

\author[Shuang Nan Zhang]   
{Shuang Nan Zhang$^{1,2}$}

\affiliation{$^1$Physics Department and Center for Astrophysics, Tsinghua University,
Beijing, 100084, China \break email: zhangsn@tsinghua.edu.cn\\[\affilskip]
$^2$Key Laboratory of Particle Astrophysics, Institute of High Energy Physics, Chinese
Academy of Sciences, Beijing, China}

\pubyear{2006}
\volume{Volume 14}  
\pagerange{119--137}
\date{?? and in revised form ??}
 \jname{Highlights of Astronomy, Volume 14} \editors{K.A. van der
Hucht, ed.}
\begin{document}

\def\aj{{AJ}}
\def\araa{{ARA\&A}}
\def\apj{{ApJ}}
\def\apjl{{ApJ}}
\def\apjs{{ApJS}}
\def\ao{{Appl.~Opt.}}
\def\apss{{Ap\&SS}}
\def\aap{{A\&A}}
\def\aapr{{A\&A~Rev.}}
\def\aaps{{A\&AS}}
\def\azh{{AZh}}
\def\baas{{BAAS}}
\def\jrasc{{JRASC}}
\def\memras{{MmRAS}}
\def\mnras{{MNRAS}}
\def\pra{{Phys.~Rev.~A}}
\def\prb{{Phys.~Rev.~B}}
\def\prc{{Phys.~Rev.~C}}
\def\prd{{Phys.~Rev.~D}}
\def\pre{{Phys.~Rev.~E}}
\def\prl{{Phys.~Rev.~Lett.}}
\def\pasp{{PASP}}
\def\pasj{{PASJ}}
\def\qjras{{QJRAS}}
\def\skytel{{S\&T}}
\def\solphys{{Sol.~Phys.}}
\def\sovast{{Soviet~Ast.}}
\def\ssr{{Space~Sci.~Rev.}}
\def\zap{{ZAp}}
\def\nat{{Nature}}
\def\iaucirc{{IAU~Circ.}}
\def\aplett{{Astrophys.~Lett.}}
\def\apspr{{Astrophys.~Space~Phys.~Res.}}
\def\bain{{Bull.~Astron.~Inst.~Netherlands}}
\def\fcp{{Fund.~Cosmic~Phys.}}
\def\gca{{Geochim.~Cosmochim.~Acta}}
\def\grl{{Geophys.~Res.~Lett.}}
\def\jcp{{J.~Chem.~Phys.}}
\def\jgr{{J.~Geophys.~Res.}}
\def\jqsrt{{J.~Quant.~Spec.~Radiat.~Transf.}}
\def\memsai{{Mem.~Soc.~Astron.~Italiana}}
\def\nphysa{{Nucl.~Phys.~A}}
\def\physrep{{Phys.~Rep.}}
\def\physscr{{Phys.~Scr}}
\def\planss{{Planet.~Space~Sci.}}
\def\procspie{{Proc.~SPIE}}
\let\astap=\aap
\let\apjlett=\apjl
\let\apjsupp=\apjs
\let\applopt=\ao
\def\phn{\phantom{0}}
\def\phd{\phantom{.}}
\def\phs{\phantom{$-$}}
\def\phm#1{\phantom{#1}}
\def\sun{\hbox{$\odot$}}
\def\earth{\hbox{$\oplus$}}
\def\lesssim{\mathrel{\hbox{\rlap{\hbox{\lower4pt\hbox{$\sim$}}}\hbox{$<$}}}}
\def\gtrsim{\mathrel{\hbox{\rlap{\hbox{\lower4pt\hbox{$\sim$}}}\hbox{$>$}}}}
\def\sq{\hbox{\rlap{$\sqcap$}$\sqcup$}}
\def\arcdeg{\hbox{$^\circ$}}
\def\arcmin{\hbox{$^\prime$}}
\def\arcsec{\hbox{$^{\prime\prime}$}}
\def\fd{\hbox{$.\!\!^{\rm d}$}}
\def\fh{\hbox{$.\!\!^{\rm h}$}}
\def\fm{\hbox{$.\!\!^{\rm m}$}}
\def\fs{\hbox{$.\!\!^{\rm s}$}}
\def\fdg{\hbox{$.\!\!^\circ$}}
\def\farcm{\hbox{$.\mkern-4mu^\prime$}}
\def\farcs{\hbox{$.\!\!^{\prime\prime}$}}
\def\fp{\hbox{$.\!\!^{\scriptscriptstyle\rm p}$}}
\def\micron{\hbox{$\mu$m}}

\maketitle

\begin{abstract}
Many similar phenomena occur in astrophysical systems with spatial and mass scales
different by many orders of magnitudes. For examples, collimated outflows are produced
from the Sun, proto-stellar systems, gamma-ray bursts, neutron star and black hole
X-ray binaries, and supermassive black holes; various kinds of flares occur from the
Sun, stellar coronae, X-ray binaries and active galactic nuclei; shocks and particle
acceleration exist in supernova remnants, gamma-ray bursts, clusters of galaxies, etc.
In this report I summarize briefly these phenomena and possible physical mechanisms
responsible for them. I emphasize the importance of using the Sun as an astrophysical
laboratory in studying these physical processes, especially the roles magnetic fields
play in them; it is quite likely that magnetic activities dominate the fundamental
physical processes in all of these systems.

As a case study, I show that X-ray lightcurves from solar flares, black hole binaries
and gamma-ray bursts exhibit a common scaling law of non-linear dynamical properties,
over a dynamical range of several orders of magnitudes in intensities, implying that
many basic X-ray emission nodes or elements are inter-connected over multi-scales. A
future high timing and imaging resolution solar X-ray instrument, aimed at isolating
and resolving the fundamental elements of solar X-ray lightcurves, may shed new lights
onto the fundamental physical mechanisms, which are common in astrophysical systems
with vastly different mass and spatial scales. Using the Sun as an astrophysical
laboratory, ``Applied Solar Astrophysics" will deepen our understanding of many
important astrophysical problems.

 \keywords{Sun: flares, Sun: corona, Sun: X-rays,
X-rays: binaries: individual (Cygnus~X-1), accretion disks, magnetic fields, gamma
rays: bursts: individual (GRB~940217), supernovae: individual (SN1987A), galaxies:
active: individual (M87), black hole physics}
\end{abstract}

\firstsection 
\section{Introduction}

To start with, I first categorize research methods in astronomy into four classes:
\begin{itemize}
\item Objects oriented: focus on studying particular types of objects, such as stars,
galaxies, or clusters of galaxies, etc. Astronomers doing this type of research are
normally called observers or observational astronomers;

\item Physical processes oriented: apply known physical processes in order to explain
all observed phenomena. Those doing such research are called theorists or theoretical
astrophysicists.

\item Data analysis/mining: apply new or existing data analysis methods to large
quantities of data, in order to discover new phenomena. We call them data analysts.

\item ``Similar phenomena at different scales": look for similar phenomena from
astrophysical systems with very different scales, in order to find common physical
mechanisms operating in them. Perhaps they can be called ``Observational Physicists".
\end{itemize}

The last class is the focus of this review article. This method may be appreciated from
another perspective. According to the well-known astronomy ladder of J.J. Drake, the
``fun" of astronomical subjects going down in this order: cosmology/black holes,
quasars/AGN, planet hunting, galaxies, Milky Way/CVs, stars and the Sun. On the other
hand, the amount or details of information astronomers can collect go almost in the
opposite order. Clearly as the Sun is a star in our backyard, it is one of the most
accessible astronomical objects. The Sun is the only object we can collect rather
direct information on its core, radiative zone, convective zone, photosphere,
chromosphere, corona and its winds. Interestingly, it has been gradually realized that
many astrophysical phenomena occurring in the Sun also take place in many other
astrophysical objects with enormously different scales. It is thus very likely that
similar physical processes are operating in all these different systems. We therefore
can use the Sun as our closest astrophysical laboratory for probing many physical
processes and then apply the knowledge to understanding other astrophysical systems.
Perhaps we can call this research discipline ``Applied Solar Astrophysics".

The remaining part of this report is divided into four sections. In section 2, I will
show some similar phenomena between the atmospheres of the Sun and that in accreting
black hole systems, between jets produced in all kinds of different astrophysical
objects with vastly different scales, and between the triple-ring structures in a young
supernova remnant and in a galaxy cluster centered at an active galaxy. Magnetic
activities seem to dominate or at least greatly influence all these phenomena. In
section 3, I will focus on a particular kind of similar phenomenon, i.e., the
non-linear dynamics of X-ray variations between that of the Sun, of a stellar mass
black hole binary and of a gamma-ray burst. Their X-ray variations show remarkably
consistent properties which can be modelled by multiplicative non-linear dynamical
processes, suggesting that their X-ray emission regions are made of many
inter-connected nodes or elements; multi-scale magnetic field topology most likely
plays a key role in the non-linear dynamics observed in all of them. In section 4, I
will make some concluding remarks on using the Sun as an astrophysical laboratory to
understand these similar phenomena at different scales, and suggest a future X-ray
instrument required to solve some existing open issues, in order to make further
progress in ``Applied Solar Astrophysics".

\section{Common magnetic activities dominating processes}

In this section, I review briefly three types of similar astrophysical phenomena over
a huge range of astrophysical scales. Observational and theoretical studies indicate
that these phenomena are all related to magnetic processes in which magnetic topology
and energy release play fundamental roles.

\subsection{Similar atmospheric structures between the Sun and black hole systems}

The temperature inside the Sun decreases from about 15 million degrees in its nucleus,
where thermal nuclear fusion produces about 99\% of the Sun's radiation energy, to
about six thousand degrees on its photosphere. However starting from its photosphere
its atmospheric temperature increases to about ten thousand degrees in its chromosphere
and to above one million degrees in its corona. Because the Sun's interior remains as
the only energy source, this temperature inversion means that the solar atmosphere is
not in an equilibrium with its interior thermal energy source. Such a temperature
inversion, i.e., the solar corona heating problem still remains one of the most
important problems in astrophysics. It is widely believed that solar flares driven by
magnetic reconnections may be able to provide the required energy source for heating
the solar corona, if sufficiently frequent `nano'-flares (about 10$^{24}$ erg energy
per event) are produced. This is the so-called Parker's conjecture
(\cite{parker83,parker88,parker91}). However up to now searches for these speculated
`nano'-flares still remain inconclusive (\cite{Walsh}). Theoretically it also remains
uncertain if it is physically possible to produce enough `nano'-flares
(\cite{Klimchuk}), even within the most hopeful avalanche models (\cite{Charbonneau}).
On the other hand, alternative models of solar corona heating may involve ion cyclotron
waves and turbulence (see, e.g., \cite{ion_heating,alfven}).

The accretion disk around a black hole in an X-ray binary system has a similar
structure to solar atmosphere (\cite{liang-price-77}) and in particular the temperature
inversion from its disk surface to its corona resembles that of the Sun, as shown in
Fig. 1 (\cite{zhang_science}). Strikingly, the temperatures of the three regions in an
accretion disk around a black hole are higher by approximately a factor of 500 than the
corresponding regions in the Sun. This supports the notion that magnetic activity is
responsible for powering the upper atmosphere in both cases, giving $T\propto
E^{1/4}\propto B^{1/2}$ and thus $T_{\rm DISK}/T_{\rm SUN}\approx (B_{\rm DISK}/B_{\rm
SUN})^{1/2}\approx (10^8 {\rm G}/500 {\rm G})^{1/2}\approx 500$, where the typical
magnetic field strengths of $~10^8 {\rm G}$ for inner regions of black hole accretion
disk (assuming equipartition) and $~$500 G for the active regions of the Sun are used.

\begin{figure}
\centerline{\hbox{\includegraphics[width=1.5in,angle=270]{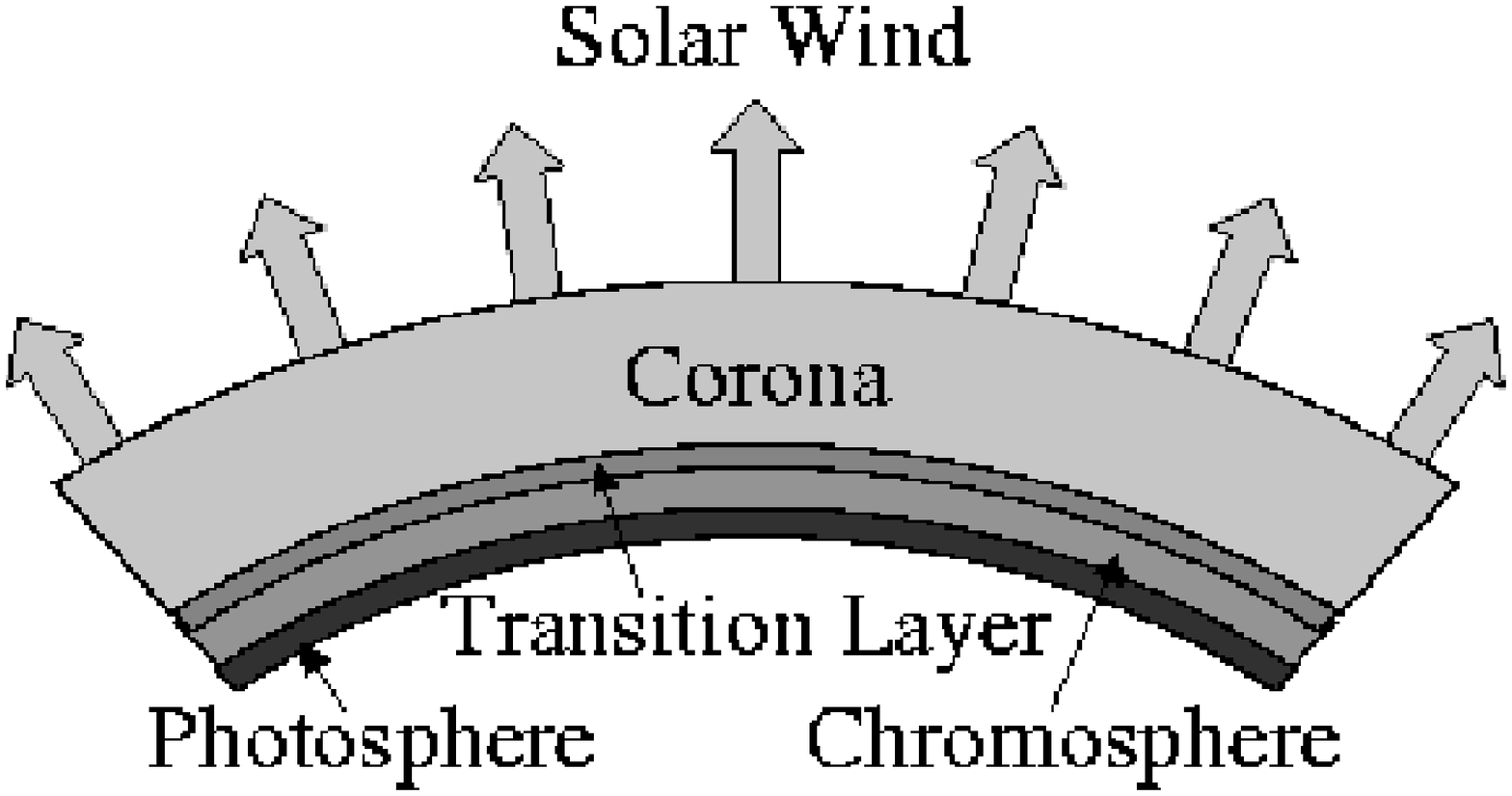}
\includegraphics[width=2.0in,angle=270]{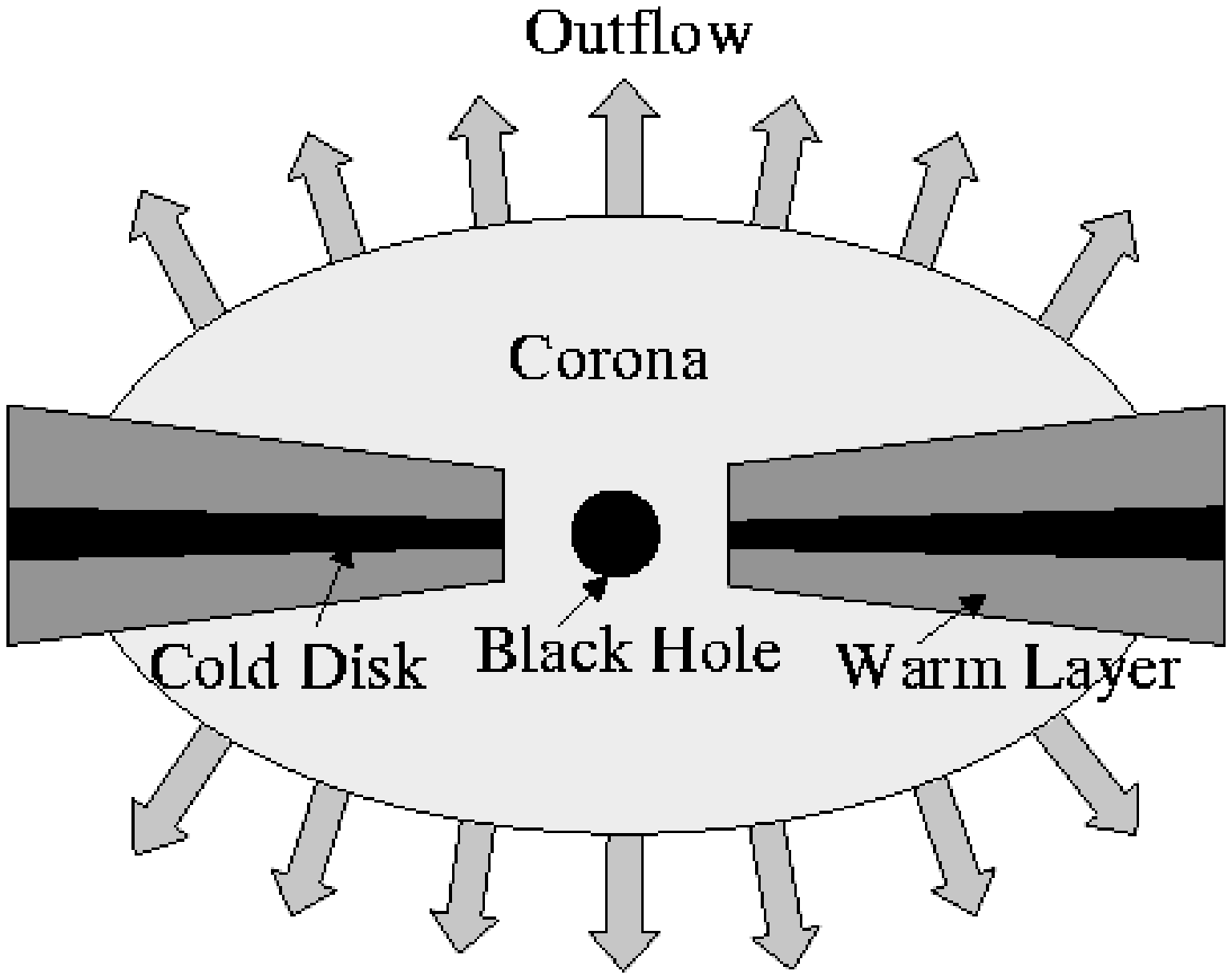} }}
 \caption{Schematic diagrams of the solar atmosphere and accretion
disk structure. The temperatures in the solar atmosphere are approximately:
6$\times$10$^3$ K (photosphere),  3$\times$10$^4$ K (chromosphere), and 2$\times$10$^6$
K (corona), respectively. For the black hole disk atmosphere, the corresponding
temperatures are approximately 500 times higher: 3$\times$10$^6$ K (cold disk),
1.5$\times$10$^7$ K (warm layer), and 1$\times$10$^9$ K (corona), respectively.
(Re-produced from \cite{zhang_science})}
\end{figure}

Since in an ionized accretion disk the angular momentum transport as well as energy
dissipation are likely dominated by the so-called magneto-rotational instability
mechanism (\cite{balbus-hawley}), X-ray flares or `shots', driven by magnetic
reconnection, may be responsible for the observed X-ray variability in systems ranging
from accreting neutron star and black hole binaries, as well as active galactic nuclei
(AGN) harboring supermassive black holes. Such a `flare' model for accretion disks
around black holes thus makes two generic predictions: (1) The X-ray emission region,
including the corona in both stellar mass and supermassive black hole systems should
have a `disk-like' geometry, rather than a spherical-like geometry, which has been
widely assumed; (2) Naturally their X-ray light curves should be made of individual
flares or `shots' if sufficiently good sensitivity and time resolution are available.

\cite{Kubota} have identified a `disk-like' configuration for the hot corona producing
a hard X-ray power-law component through inverse Comptonization process, in the stellar
mass black hole binary GRO~J1655-40, which was discovered by \cite{iauc_1655} and in
fact the second microquasar (X-ray binaries with superluminal jets) in the Milky Way
(\cite{Tingay,Harmon}. The X-ray emitting coronae in AGNs are suggested to have a
`disk-like' geometry, in order to explain the decrease of fraction of Type II AGNs as a
function of the observed apparent X-ray luminosity (e.g., \cite{ueda}), due to the
smaller projected area of the `disk-like' corona for Type II AGNs
(\cite{zhang_agn,liu_zhang}). Flares or `shots' are commonly observed from accreting
systems of different sizes
(\cite{terrell,doi,Miyamoto_88,Haardt,Nayakshin,Matteo,Poutanen,Negoro01,wang,Liu04}).
Modelling of periodic or quasi-periodic X-ray and infrared flares from Sgr A*, the
supermassive black hole in the center of the Milky Way, suggests that these flares are
of accretion disk origin (\cite{aschenbach}). It is thus quite possible that these
flares are also produced by processes driven by magnetic reconnections.

We note that a hard X-ray power-law component up to at least 50-100 keV have also been
seen from from active galactic nuclei (\cite{zhang_bassani}) and weakly magnetized
(with surface magnetic field strength of around $~10^8 {\rm G}$) neutron star X-ray
binaries (e.g., {\cite{zhang_1608,zhang_iau,chen_zhang}), which often exhibit similar
spectral state transitions as black hole binaries ({\cite{zhang_iau,zhang_cygx1}). It
is well understood that the standard optically thick and geometrically thin accretion
disk models (\cite{diskmodel}) cannot produce this power-law component. However the
orbital kinetic energies of protons in the inner accretion disk regions in weakly
magnetized neutron star X-ray binaries, stellar mass black hole binaries and
supermassive black hole systems should be approximately the same. Therefore if the
energy source of the above mentioned `flare' model for accretion disks is dominated by
particle's kinetic energies, the similar hard X-ray power-law component may be
generated naturally. We comment in passing that the solar magnetic fields are also
believed to be generated by dynamo mechanisms due to differential rotation in the Sun.

\subsection{Astrophysical jets at different scales}

Collimated outflows or jets are common astrophysical phenomena, now found in the Sun,
proto-stellar systems, isolated neutron stars, neutron star and black hole binaries,
gamma-ray bursts and supermassive black holes. MHD simulations have shown that
differential rotation and twisted magnetic fields are two generic ingredients for
generating collimated outflows (\cite{meier}), somewhat similar to magnetic
reconnection processes in which twist and shear of magnetic flux tubes play very
important roles for solar flares and perhaps also coronal mass ejection events (see,
e.g., {\cite{wang_cme,Pevtsov,helicity}). With these two generic ingredients, their
different scales and observational appearances (such as Lorentz factor and degree of
collimation) of different astrophysical jets may reflect their different astrophysical
environments (such as gravitational potential, differential rotation energy, magnetic
field strength and topology, spin of the central object, etc), where jets are produced,
accelerated, collimated and transported. Therefore astrophysical jets can be used as
powerful astrophysical probes for many different types of objects. For example,
prolonged optical and X-ray afterglow emissions from gamma-ray bursts have been used to
probe the properties of interstellar media around the gamma-ray burst progenitors and
consequently the nature of their progenitors, because gamma-ray burst afterglow
emissions are believed to be produced from external shocks when their jets eventually
plow into their surrounding media, in a similar way to an expanding supernova shell
plowing into its surrounding medium. Interestingly the same external shock model has
been used to explain the decelerated relativistic jet from a black hole binary system
 (\cite{wang_jet}), implying a dense surrounding medium and thus a massive
star as the progenitor of the stellar mass black hole, similar to the progenitors to
many long duration gamma-ray bursts (see, e.g., \cite{mesaros} for a review and
references therein).

In many cases jets from black hole binaries, gamma-ray bursts and supermassive black
holes are relativistic and may share the same production mechanisms, and it is very
likely that black hole's spin energy is extracted to power their relativistic jets
(see, e.g., \cite{zhang_gro_review,mirabel,mirabel_sky} for reviews and references
there in). Using the X-ray continuum spectra, Zhang, Cui \& Chen (1997) first
determined the spin of the black holes in several black hole binaries with relativistic
jets and established the connection between black hole spin and jet observationally.
This method of determining black hole's spin has since been further refined and widely
applied to many more black hole binaries (for a recent review, see, e.g.,
\cite{remillard_araa}); the existence of extremally spinning black hole is now
well-established.

\subsection{Similar triple-ring structures between a young supernova remnant and a
galaxy cluster}

\begin{figure}
\centerline{\hbox{\includegraphics[width=2.5in]{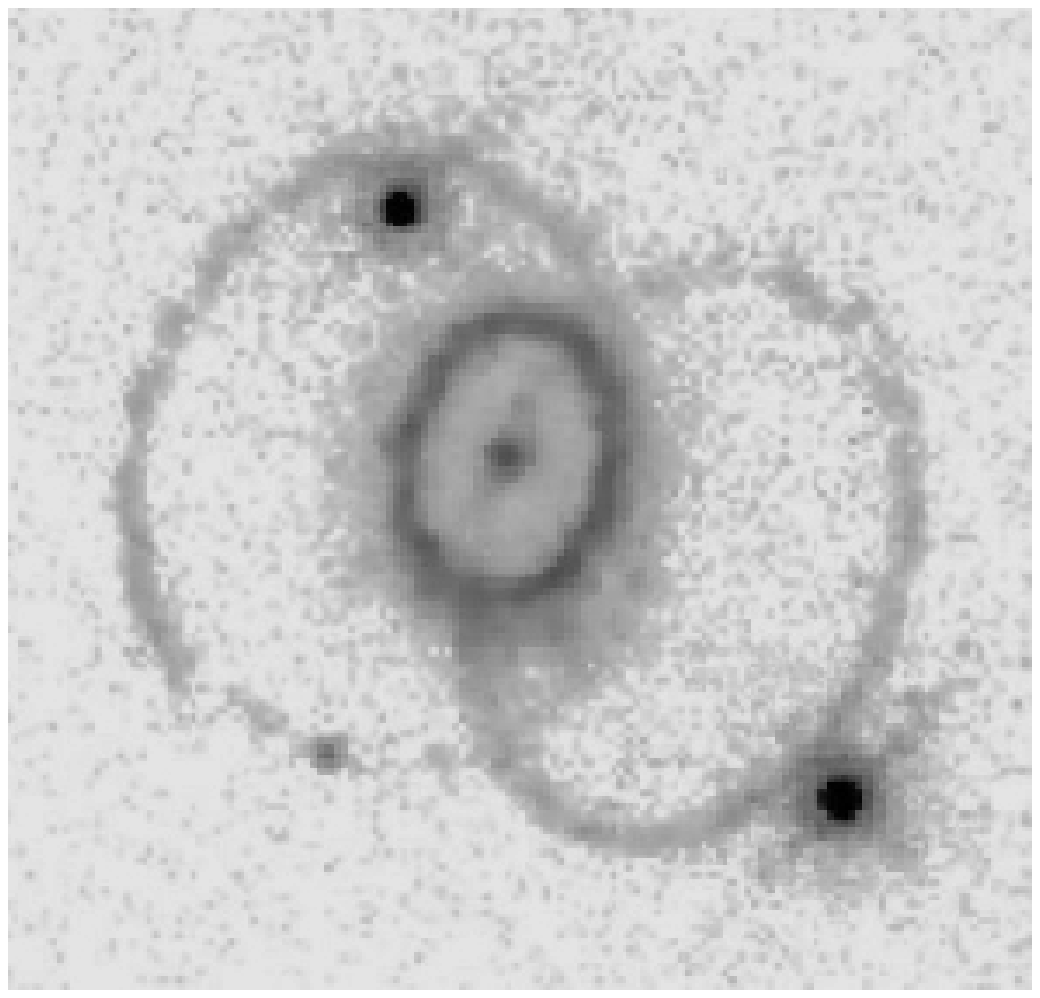}
\includegraphics[width=2.5in]{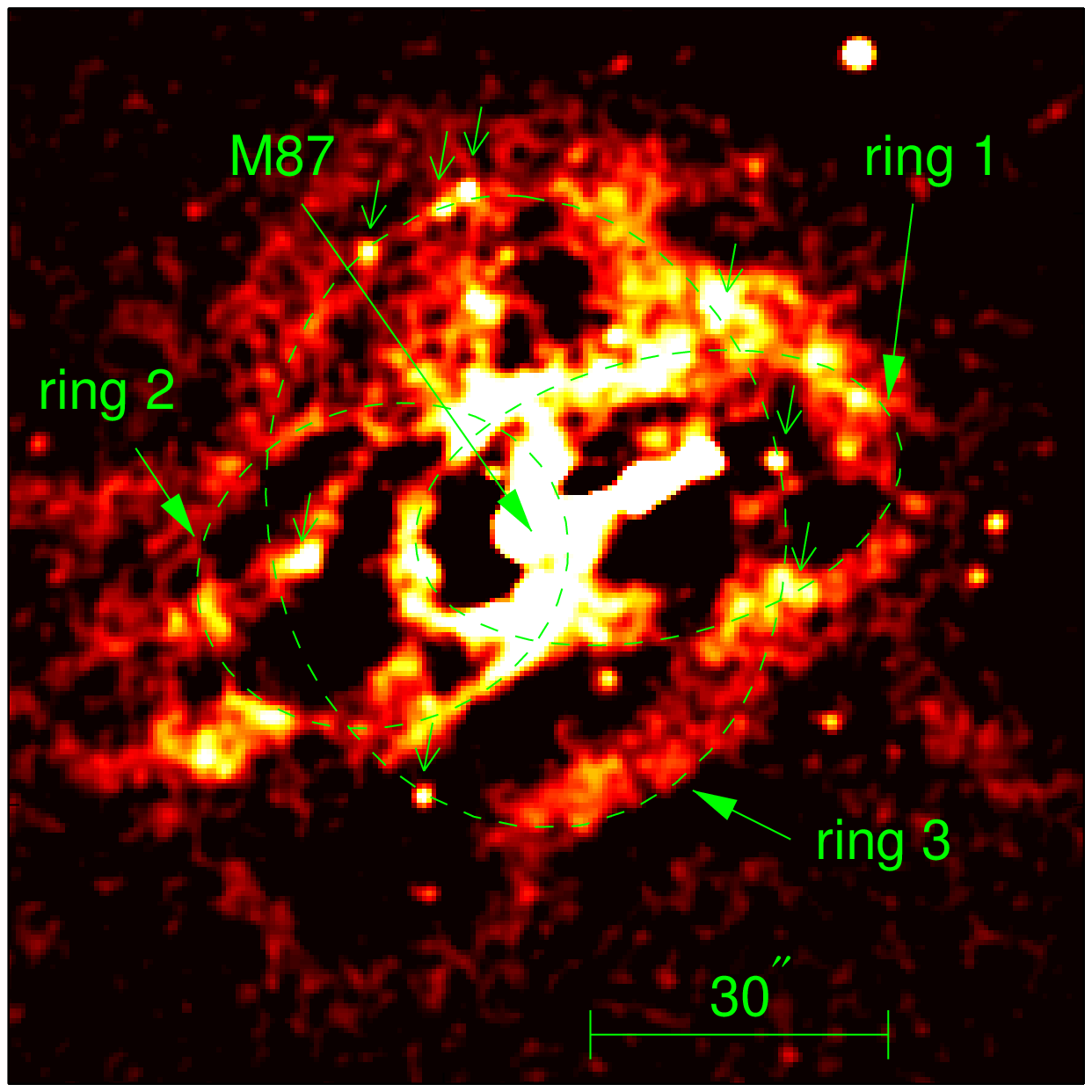}}}
 \caption{Similar triple-ring structures between the supernova remnant of SN1987A and
Virgo galaxy cluster centered at the active galaxy M87. {\it Left}: The optical
triple-ring structure of SN1987A, as observed with Hubble Space Telescope
(\cite{sn1987a}). {\it Right}: The X-ray triple-ring structure of Virgo galaxy cluster
as observed with the Chandra X-ray observatory (\cite{feng}).}
\end{figure}

Another intriguing pattern of similarity is between the supernova remnant of SN1987A
and M87, an active galaxy producing a large scale jet in the center of the Virgo
Cluster (Fig.~2). The triple-ring structure of SN1987A observed in visible band with
the Hubble Space Telescope is perhaps the most beautiful astronomical picture taken so
far (see Fig. 2 (left); \cite{sn1987a}). A MHD simulation by Tanaka \& Washimi (2002)
showed that if its progenitor has similar large-scale magnetic field structure like the
Sun, then its winds, similar to solar winds, will follow the twisted magnetic field
topology and are thus preferentially located along its spin axis or in the equatorial
plane. Suppose its progenitor has experienced a slow-wind phase (red-giant) and then a
fast-wind phase (blue-giant), then the fast winds will eventually catch the slow winds
to form the triple-ring like structure. It is possible that a triple-ring structure
might be generic in a rotating system involving magnetized faster winds catching up
slower winds. Meanwhile a central source of radiation is required for the rings to
shine. For SN1987A, the strong UV radiation from the supernova provides the necessary
illumination.

In Fig.~2 (right) a triple-ring like structure in the X-ray band similar to that of
SN1998A was shown, observed with the Chandra X-ray observatory (\cite{feng}). The ring
sizes in the Virgo cluster are several thousand times larger than those of SN1987A. It
is possible that a catastrophic merging event around the M87 nucleus is responsible for
the triple-ring structure revealed here. In this scenario, a ``slower wind'' was
present before the merging begins, e.g., a mixture of galactic winds from two merging
galaxies. A ``faster wind'' was then driven during the merging process. The final
merging of the two supermassive black holes releases a huge amount of high energy
radiation to shine the triple rings in X-ray bands. Presumably, the resulting
supermassive black hole spins rapidly to power the highly collimated M87 jets.

\section{Non-linear dynamical processes at different scales}
\begin{figure}\label{lc_2022619}
\centerline{\includegraphics[width=5.0in]{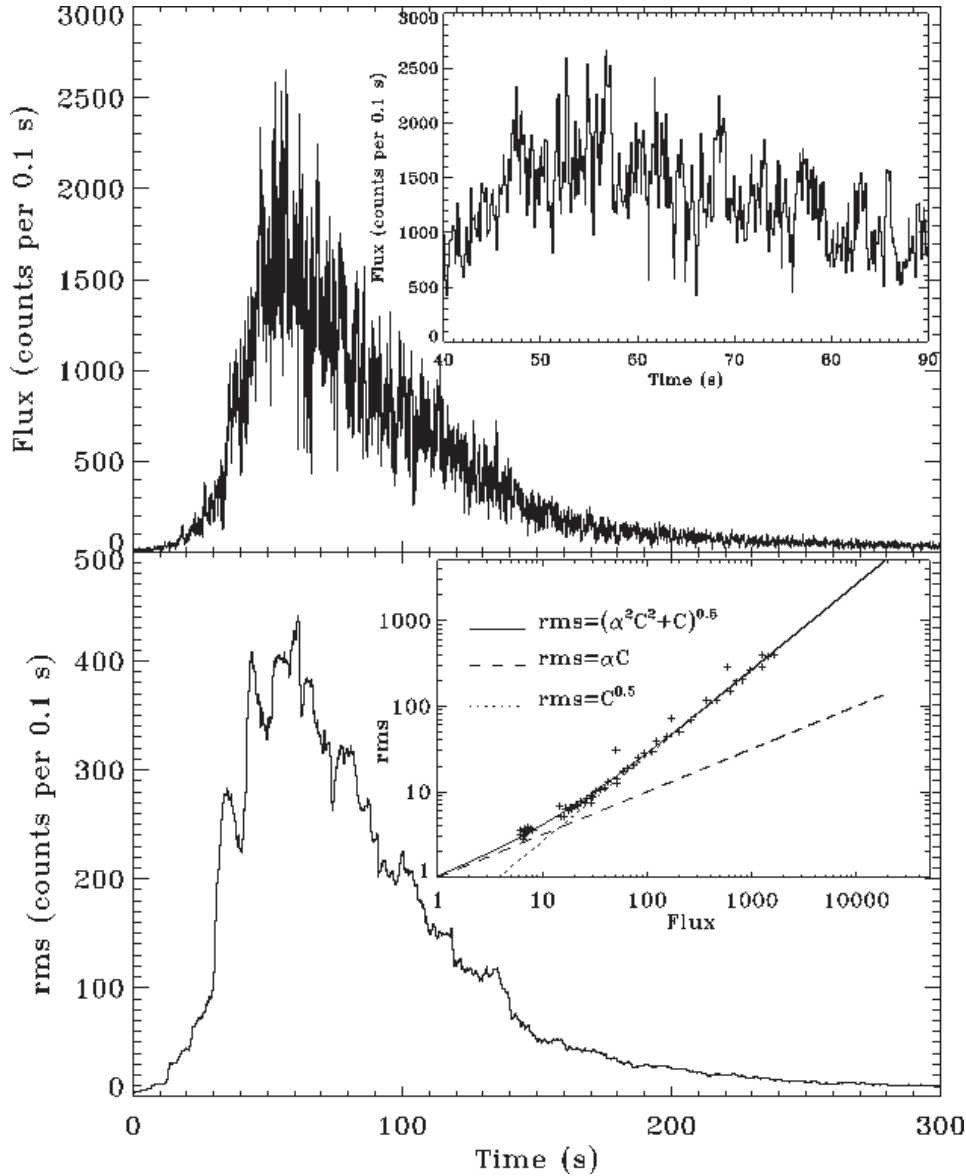}} \caption {{\it RHESS}
2022619 solar X-ray flare between 3-50 keV. The upper panel and inset show its
lightcurve at 0.1 second resolution; variability beyond Poisson counting fluctuations
is clearly visible. The lower panel shows the rms (root-mean-squares), in units of
counts in 0.1 second, for each 1 second internal as a function of time and the average
flux, in units of counts per 0.1 second, during each 1 second interval. For the model
lines shown in the inset, $\alpha=0.27$ is the best fit result with negligible error.}
\end{figure}

\begin{figure}\label{all_flares}
\centerline{ \includegraphics[width=5.0in]{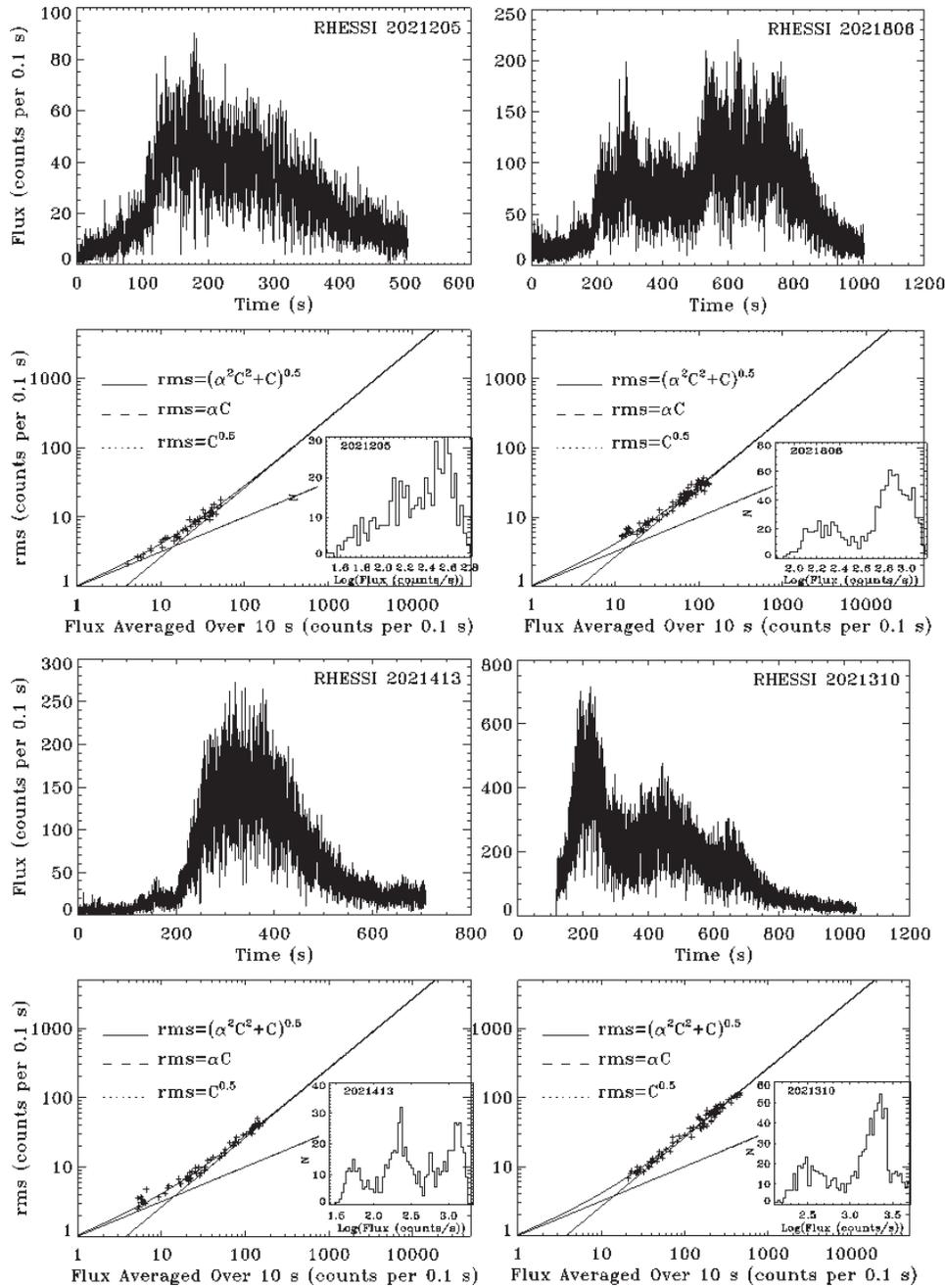}} \caption {The
lightcurves, flux-rms correlation and distribution of the logarithm of the flux for
four individual solar X-ray flares. It should be noted $\alpha=0.27$ for all of them,
and that their distributions of the logarithm of the flux are made of one or several
normal distributions.}

\end{figure}
\begin{figure}\label{all_flares}
\centerline{ \includegraphics[width=5.0in]{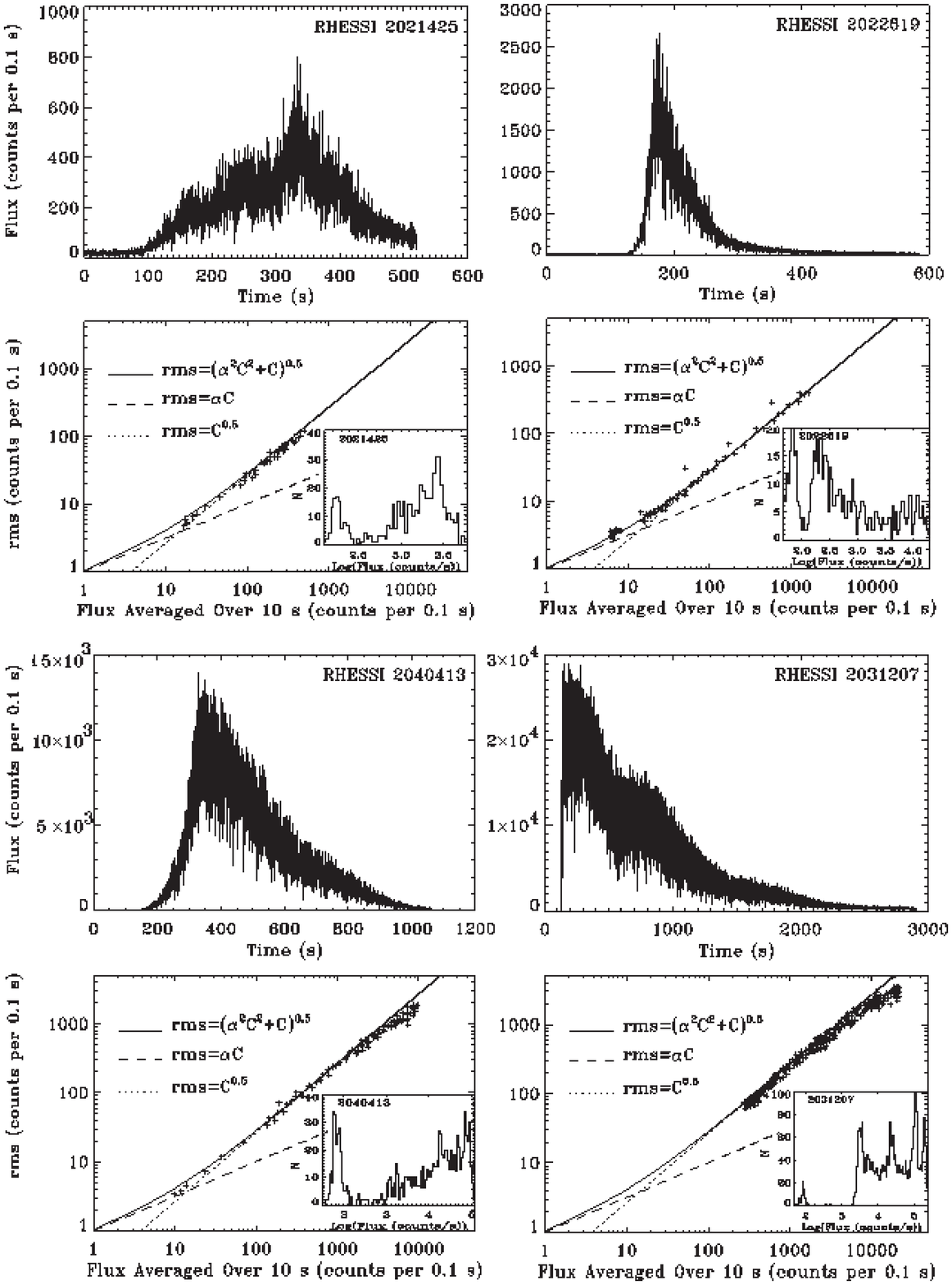}} \caption
{Continuation of Fig.~4.}

\end{figure}

 In an extensive study of X-ray variability
from a prototype black hole X-ray binary Cygnus~X-1 (\cite{cygx-1-1}), \cite{Uttley05}
have shown that its X-ray flux fluctuations, characterized by rms (root-mean-squares)
of its flux, is proportional to its mean flux. Such a flux-rms linearity cannot be
modelled by additive or deterministic processes. Instead, \cite{Uttley05} have
demonstrated that the dynamical process for the X-ray production should be a non-linear
`multiplicative' process, which can be modelled by the following non-linear time-series
model,
\begin{eqnarray}
\label{voltnew} X_{i} & = & 1+\sum^{\infty}_{j=0} G_{j}u_{i-j}
+\sum^{\infty}_{j=0}\sum^{\infty}_{k=0}G_{jk}u_{i-j}u_{i-k} +
\nonumber \\
& & \sum^{\infty}_{j=0}\sum^{\infty}_{k=0}\sum^{\infty}_{l=0}
G_{jkl}u_{i-j}u_{i-k}u_{i-l}+ ...
\end{eqnarray}
where the coefficients $G_{j}, G_{jk}, G_{jkl}\ldots$ and the higher-order
co-efficients are non-zero. Therefore the flux-rms linearity is a clear indication of
non-linear dynamical process. Combining this and the fact that the flux distribution of
Cygnus~X-1 follows a log-normal distribution, but the peak flux distribution of solar
flares follows approximately a power-law distribution (which is predicted by the
self-organized criticality model), \cite{Uttley05} thus claimed to have rejected
essentially all previous models for X-ray variability from X-ray binaries and active
galactic nuclei, including shot noise, self-organized criticality, or dynamical chaos
models (\cite{Uttley05}).

Since the characteristic X-ray variability seen in Cygnus~X-1 has also been seen from
other black hole X-ray binaries (\cite{Uttley01}), as well as neutron star
(\cite{Uttley04}) and supermassive black hole accreting systems
(\cite{edelson,Vaughan_a,Vaughan_b}), it has been suggested by \cite{Uttley05} that the
X-ray variability in these systems should be driven by fluctuating accretion flow
(\cite{Kotov,Lyubarskii}). Since accretion flow does not exist in the Sun, their
conclusion rules out common physical mechanism for X-ray emission from the Sun, thus
has far-reaching impacts to our understanding of some fundamental processes in a very
broad range of astrophysical systems.

Motivated by this and in light of the outstanding solar corona heating problem, we
examine the X-ray variability in the Sun's X-ray lightcurves collected by the Reuven
Ramaty High Energy Solar Spectroscopic Imager ({\it RHESSI})
(\cite{lin02})\footnote{http://hesperia.gsfc.nasa.gov/hessi/}, which has been in
productive operation since launch on 5 February 2002 (\cite{dennis}). We also
re-examine the non-linear dynamical properties of Cygnus~X-1 and a gamma-ray burst. We
demonstrate that a simple model can describe the non-linear dynamical properties of
these three kinds of systems of very different astrophysical objects at very different
scales.

\subsection{flux-rms relation and flux distributions for individual solar X-ray flares}
\begin{figure}\label{lc_jan}
\centerline{\hbox{\includegraphics[width=2.6in]{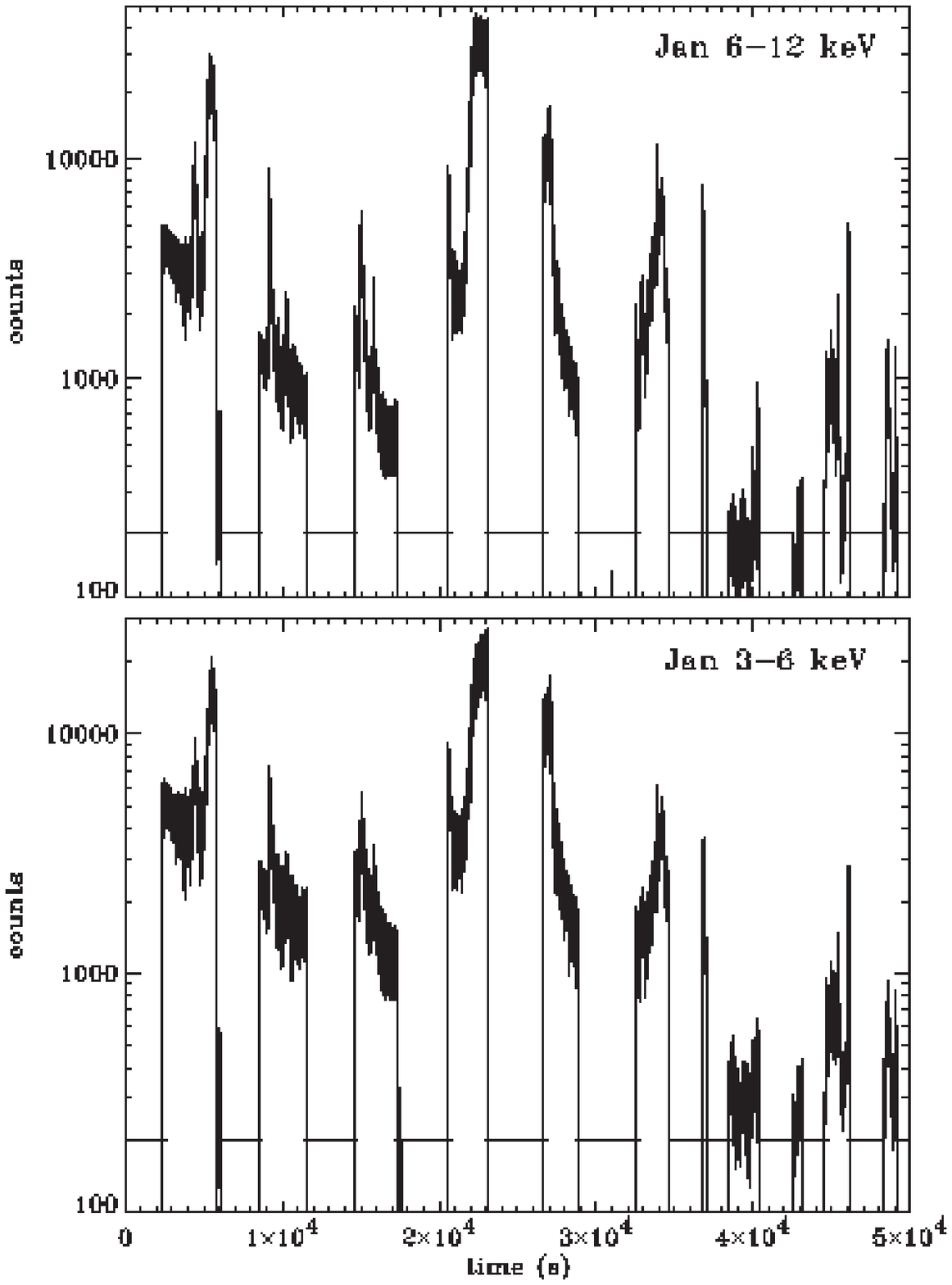}
\includegraphics[width=2.6in]{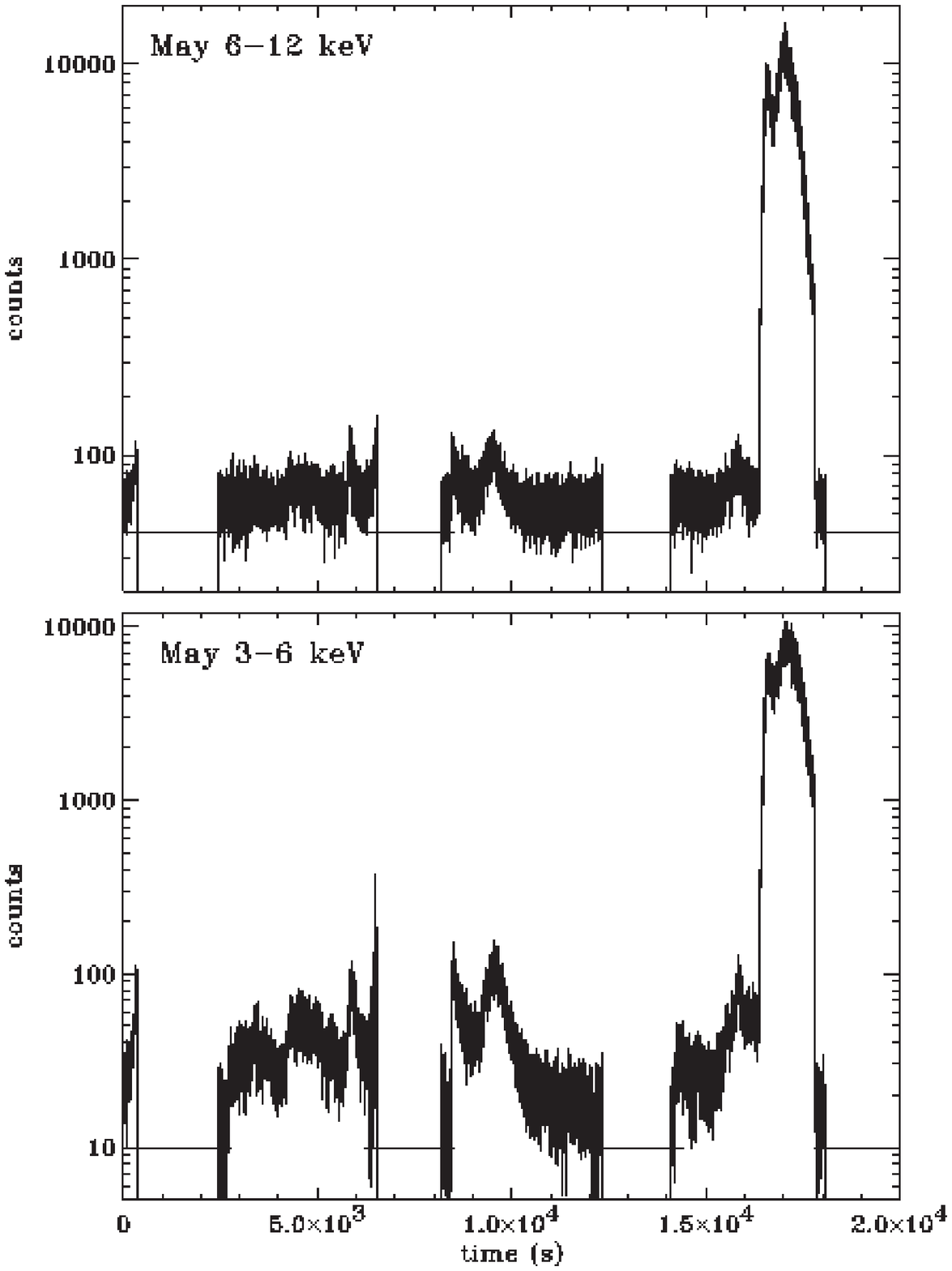}}}
 \caption {Solar X-ray lightcurves in two energy bands with 1 s resolution, on
2004-Jan-1.0 (left panels) for 50,000 consecutive seconds and on 2004-May-1.0 (right
panels) for 20,000 consecutive seconds. Horizontal straight bars near the bottom in
each panel mark data gaps, i.e., `night' times for {\it RHESSI} satellite.}

\end{figure}

\begin{figure}\label{lc_jan}
\centerline{\hbox{\includegraphics[width=2.6in]{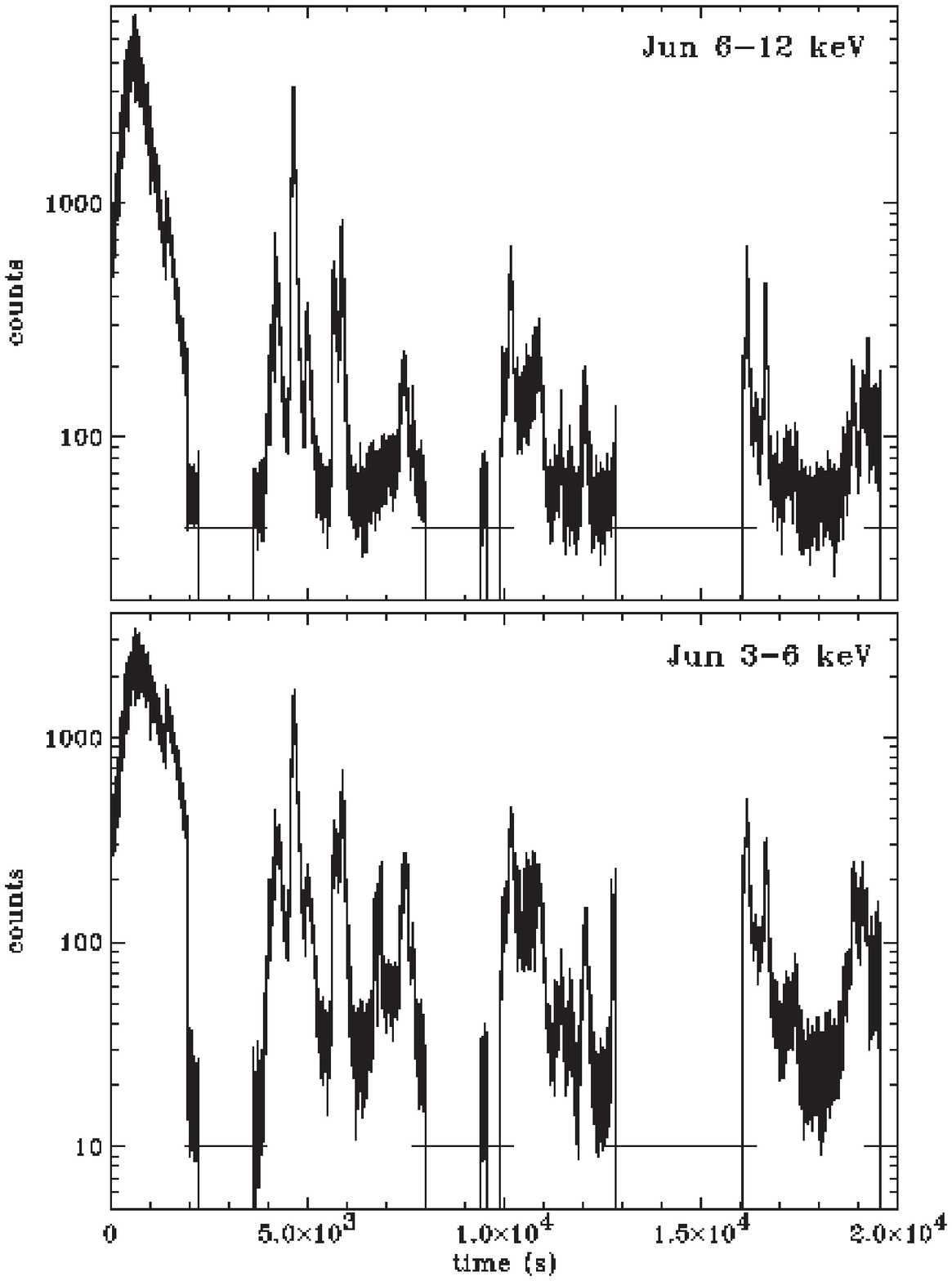}
\includegraphics[width=2.6in]{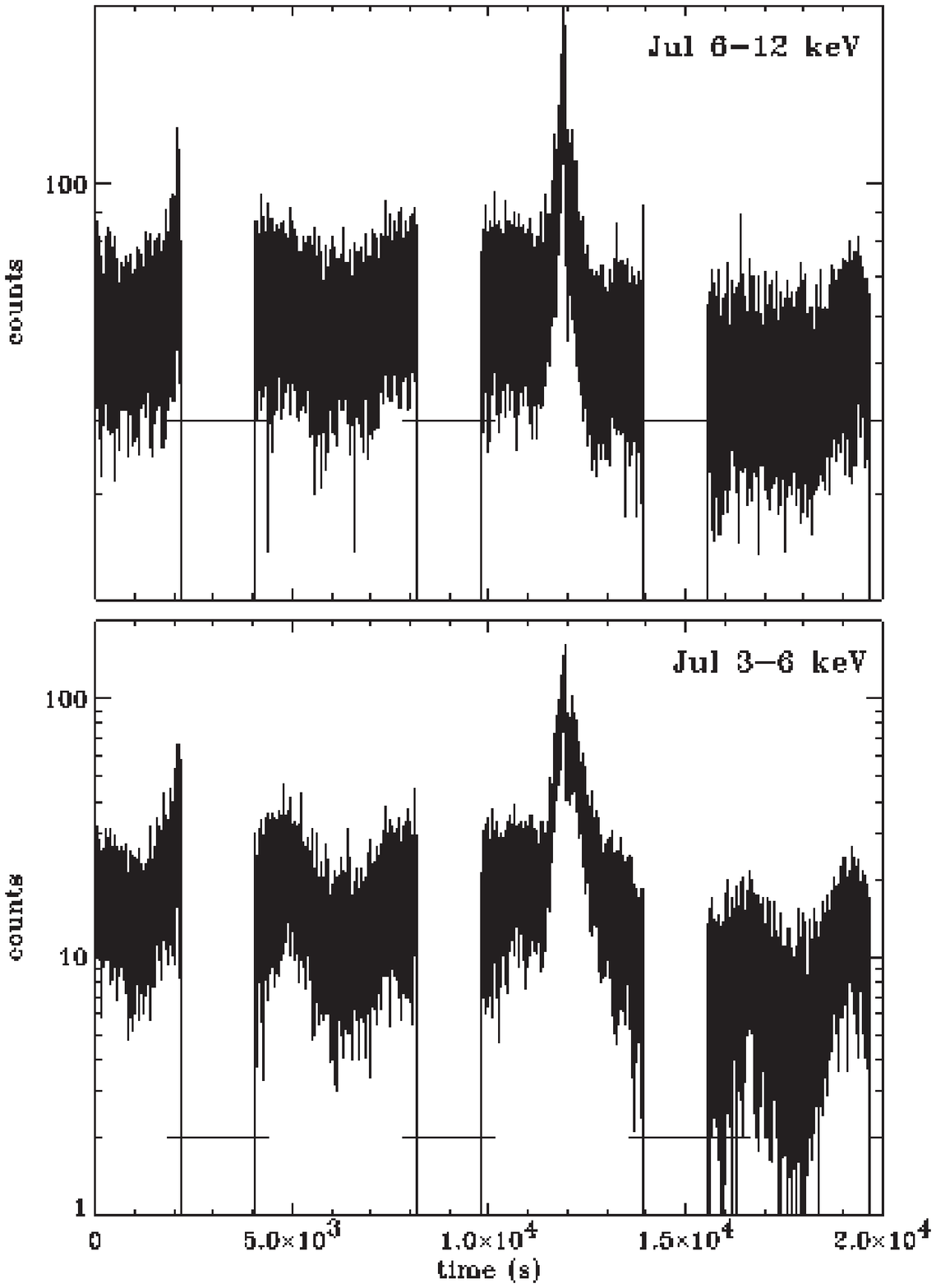}}}
 \caption {Solar X-ray lightcurves in two energy bands with 1 s resolution, on
 2004-Jun-1.0(left panels) for 20,000 consecutive seconds and on 2004-Jul-1.0 (right
panels) for 20,000 consecutive seconds. Horizontal straight bars near the bottom in
each panel mark data gaps, i.e., `night' times for {\it RHESSI} satellite.}

\end{figure}

\begin{figure}\label{sun_flux_rms}
\centerline{ \includegraphics[width=5.2in]{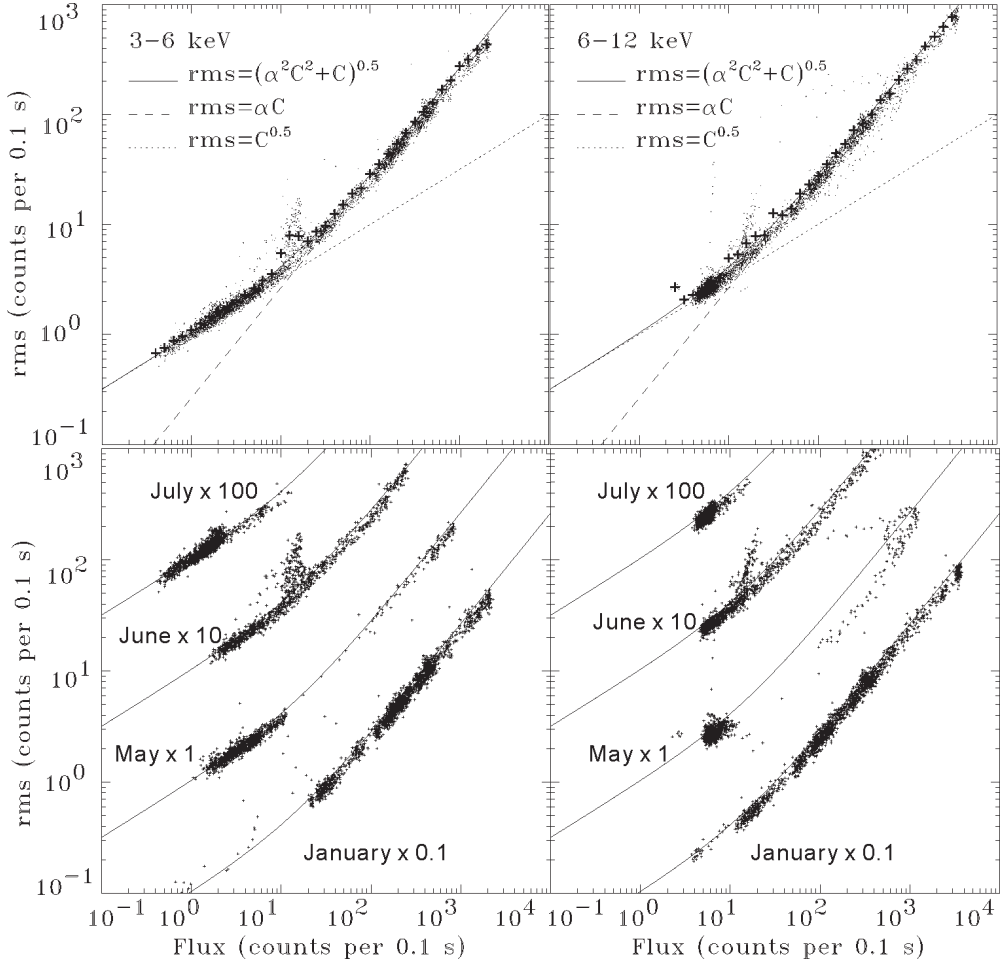}
} \caption {{\it Upper panels}: The relationship between flux and rms  for all four
segments of lightcurves in two energy bands. Each point in the scatter plots is for a
10 s interval of data with 0.1 s resolution. {\it Lower panels}: The flux-rms relations
for the four lightcurves are shown separately; the data and model lines for January,
June and July are shifted vertically for clarity. Clearly all four lightcurves have
consistent flux-rms relations.}

\end{figure}

\begin{figure}\label{cyg_flux_rms}
\centerline{ \includegraphics[width=3.5in]{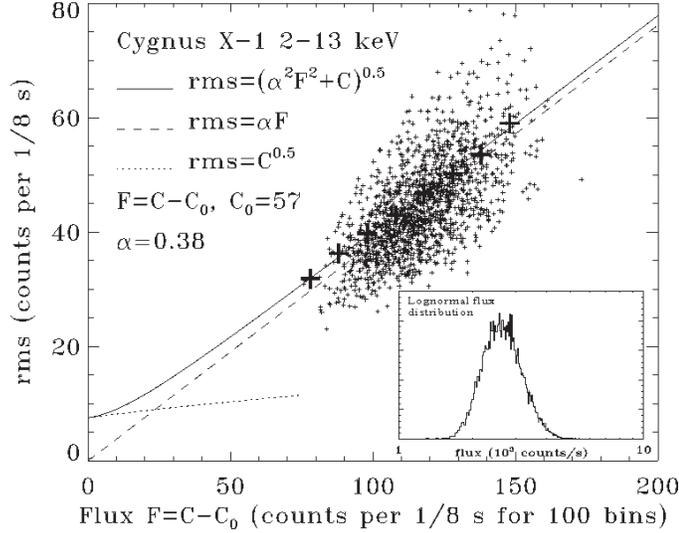} } \caption {The
flux-rms relation for the canonical stellar mass black hole binary Cygnus~X-1 in the
2-13 keV, with data collected with the PCA instrument onboard the Rossi X-ray Timing
Explorer from 1996-10-23-18:30:24 to 1996-10-24-02:30:26. The crosses for the average
values of rms for each flux interval divided uniformly in linear scale due to the
narrow flux range. Note counts are for per 1/8 s time bin; each point in the scatter
plots is for a 10 s interval of data with 1/8 s resolution. The inset shows its
log-normal flux distribution.}

\end{figure}

\begin{figure}\label{grb1}
\centerline{\includegraphics[width=3.5in]{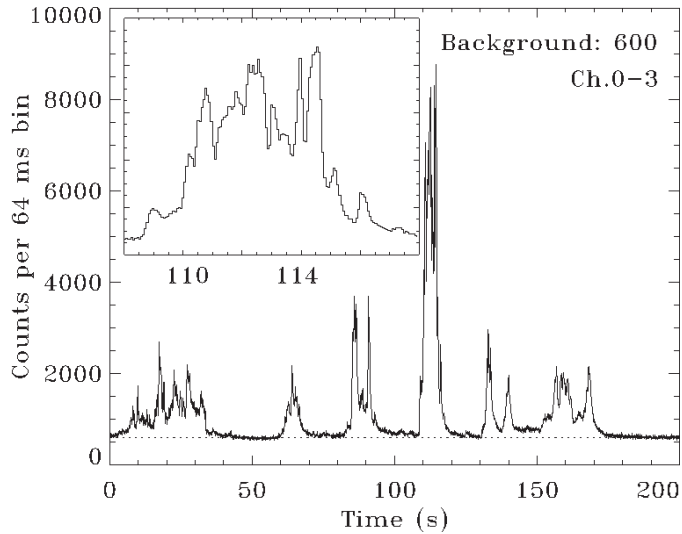}
} \caption {The lightcurve of GRB~940217 (trigger No. 2831) in channels 0-3 for the 16
channel data, i.e., between 13 to 54 keV, collected with the BATSE instrument onboard the
Compton Gamma-ray Observatory. The inset shows clear variability above Poisson
fluctuations. Note the background level is about 600 per 64 ms bin.}

\end{figure}

\begin{figure}\label{grb2}
\centerline{\includegraphics[width=3.5in]{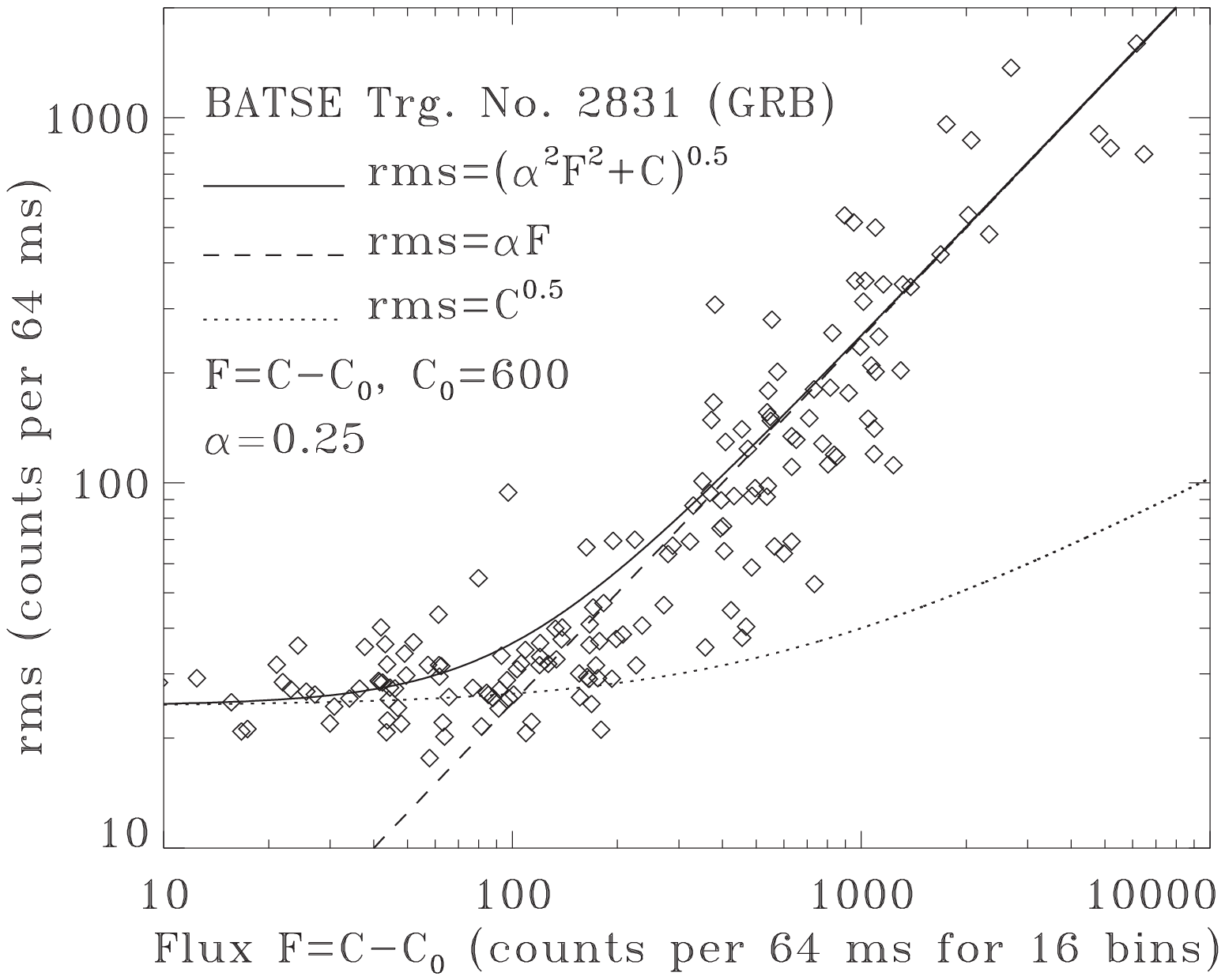}
} \caption {The flux-rms relation for the lightcurve of GRB~940217. The model is the
same as equation (3.3).}

\end{figure}

In Fig.~3 the lightcurve, rms evolution and flux-rms correlation for {\it RHESSI}
2022619 solar flare are shown. The flux-rms relationship shown in the inset of Fig.~3
can be described by a simple
 model (plotted as solid line),
\begin{equation}
{\rm rms}^2=(\alpha C)^2+C,
\end{equation}
where $\alpha=0.27$ is a constant, $C$ is the average counts within each 1 second
interval. Clearly the total rms observed is made of two terms, a term linearly
proportional to the average flux of the solar flux at a given time (plotted as dashed
line), and a Poisson fluctuation term (plotted as dotted line); for a random Poisson
process, rms~=~$\sqrt{C}$. It can be seen from Fig.~3 that the linear term starts to
dominate at flux above 20 counts per 0.1 second.

In Fig.~4 and Fig.~5, eight {\it RHESSI} solar flares with different peak fluxes,
durations and lightcurve morphologies are shown to illustrate their common properties.
Interestingly, not only can the same simple model in equation (3.2) describes their
flux-rms relation perfectly, but also $\alpha=0.27$ for all of them. In the insets the
flux distribution of each solar flare is also plotted, which can be approximated by one
or several log-normal components, i.e., the logarithm of the flux follows normal
distributions, in comparison for the single log-normal peak of flux distribution for
the black hole X-ray binary Cygnus~X-1 (\cite{Uttley05}), and see the inset of Fig.~8
in section~3.3).

\subsection{flux-rms relation for continuous solar X-ray light curves}

In Figs.~6 \& 7 several continuous segments of characteristic {\it RHESSI} solar X-ray
lightcurves are shown in 3-6 keV and 6-12 keV bands with 1 second time resolution.
These figures reveal X-ray variability at all time scales with a large range of
amplitudes over about five orders of magnitudes. Despite of its apparently very
different lightcurves observed at well separated times, the solar X-ray variability
exhibits remarkably consistent and simple relationship between the rms
 and its mean flux as shown in Fig.~8, which are again modelled by equation (3.2) with the same value of $\alpha=0.27$.
 We note that the
flux-rms linearity also exists in a very broad frequency range if calculated from the
FFT power-spectra of its lightcurves (\cite{tang_zhang}), again similar to that of
Cygnux~X-1 (\cite{Uttley05}). It should be noted that a continuous solar X-ray
lightcurve is collected over the whole surface
 of the Sun, not just from a single active region of the solar surface as for the case of a single
 solar flare. According to the mathematical description of equation (3.1) for the linear flux-rms relationship, this implies that many active regions in the solar surface are not
 isolated, but somehow connected and react to common triggers or ignitions. This is because the flux fluctuations from many isolated or independent regions should be proportional to the square-root
 of the total combined flux. Such inferred inter-connection between different
 active regions may offer deep insight for the solar activity mechanisms.

\subsection{Cygnus~X-1 and a gamma-ray burst}

 In Fig.~9
the flux-rms correlation for the canonical stellar mass black hole binary Cygnus~X-1,
observed with the PCA instrument onboard the Rossi X-ray Timing
Explorer\footnote{http://astrophysics.gsfc.nasa.gov/xrays/programs/rxte/pca/}, is shown
(see \cite{Uttley05} for comprehensive results on the nonlinear behaviours of
Cygnus~X-1 with more data). The crosses are for the average values of rms for each flux
interval divided uniformly in linear scale due to the narrow flux range. Compared to
the flux-rms relation for {\it RHESSI} solar X-ray flares, the model describing the
correlation is only slightly more complicated,
\begin{equation}
{\rm rms}^2=(\alpha F)^2+C,\\\
F=C-C_0,
\end{equation}
where the total count $C$ consists of a constant term $C_0=57$ and a variable flux with
rms~=~0.38$\times F$ over the Poisson counting fluctuations. Note that $C_0=57$ is well
above the PCA background level, and is thus an intrinsic component from Cygnus~X-1. The
inset in Fig.~9 shows the log-normal flux distribution of Cygnus~X-1. It has been known
that similar linear flux-rms relation exists for neutron star (\cite{Uttley04}) and
black hole binaries (\cite{Uttley01}), as well as for supermassive black hole binaries
(\cite{edelson,Vaughan_a,Vaughan_b}).

In Figs.~10 \& 11 the lightcurve and flux-rms relation for GRB~940217, detected with
the BATSE instrument onboard the Compton Gamma-ray
Observatory\footnote{http://cossc.gsfc.nasa.gov/docs/cgro/batse/}, are shown for
comparison. Equation (3.3) provides a consistent description for the flux-rms relation
with $\alpha=0.25$ and $C_0=600$. Compared to the flux-rms relation for Cygnus~X-1,
$C_0=600$ is entirely due to the background level of the BATSE instrument. Therefore
the flux-rms relation for GRB~940217 is described by exactly the same model for {\it
RHESSI} solar X-ray flares, if the different instrumental background levels are taken
into account.

We chose to use GRB~940217 as an example of gamma-ray bursts, because of its large
dynamical range of flux and very long duration (thus many photons are available for a
statistically meaningful correlation for a single gamma-ray burst), as well as its very
complex light curve morphology, for illustrating the robustness of the correlation. We
comment in passing that GRB~940217 is a very important gamma-ray burst in its own
right, because of its long-lasting high energy afterglow emission detected; $>$30 MeV
photons were recorded for about 5400 seconds, including an 18 GeV photon about 4500s
after the low energy gamma-ray emission had ended in the BATSE band as shown in Fig.~10
(\cite{hurley}). However, despite of the similar rms-flux correlation between
GRB~940217 and solar flares, it is not meaningful to include multiple gamma-ray bursts
in one correlation plot, because different gamma-ray bursts are expected to have quite
different characteristic time scales, since it is now known that long duration
gamma-ray bursts have a large range of relativistic beaming factors and are produced in
the whole Universe up to at least the redshift of 10 (see, e.g.,
\cite{lin_zhang_grb,zhang_bin}).

\subsection{Further comments on the flux-rms correlations}

For the X-ray flux-rms correlations from the Sun, Cygnus~X-1, and GRB~940217, about the
only qualitative differences are: (a) There is a non-negligible constant flux from the
total X-ray flux of Cygnus~X-1 but not from the Sun or GRB~940217; (b) The range of rms
variations from the Sun and GRB~940217 covers over several orders of magnitudes, but
only about a factor of two for Cygnus~X-1; (c) The flux distribution for each solar
flare is made of one or more log-normal peaks, in comparison for only a single
log-normal peak for that of Cygnus~X-1. For the Sun and GRB~940217, even the portion of
Poisson noise domination is seen clearly. It is commonly known that solar X-rays are
produced from flare regions on solar surface, prompt X/gamma-rays from gamma-ray bursts
(at least for long duration gamma-ray bursts such as GRB~940217) originate from the
colliding relativistic jets (the so-called internal shock model, see, e.g.,
\cite{Piran} for a recent review and references therein), and X-rays from black hole
binaries are emitted from the accretion disks around their central black holes. The
remarkable common X-ray flux variations between the apparently very different systems
at different scales suggest that there should be a common and dominating mechanism
operating in all of them, although we cannot exclude the possibility that different
physical mechanisms may generate very similar X-ray flux variations. Because the Sun is
our closest astrophysical laboratory from which we can obtain a wealth of information,
in particular the direct measurements of its magnetic field topology and activities,
understanding the solar X-ray production mechanism may be a key to revealing the black
hole accretion disks and gamma-ray burst jets.

The success of this simple model with only one parameter in describing the very complex
solar X-ray flares and continuous lightcurves, observed at well separated times in two
energy bands, implies that the linear flux-rms relationship is very helpful for our
understanding of X-ray emission from the Sun, and perhaps also the corona heating
problem. Previously EUV and X-ray variability in the pixels of focal plane detectors of
{\it TRACE} and {\it YOHKOH} have been analysed in order to probe the contribution of
the speculated `nano'-flares to the solar corona heating
(\cite{Shimizu,Krucker,Aschwanden_I,Aschwanden_II,Parnell,Katsukawa01,Benz,Katsukawa03}).
However it is still not settled if the amount of `nano'-flares can fulfill the
requirement of Parker's conjecture (\cite{Walsh}). Nevertheless these studies have
indicated that probing X-ray variability is probably so far the best way for studying
this outstanding problem in astrophysics. However due to the very limited counting
statistics in those previous studies, no individual counting excess above noise
fluctuation can be identified as real flaring events. It is thus uncertain if these
variability studies have revealed unambiguously `nano-'flares as extrapolated smaller
events from observed solar flares.

It is interesting to note that the detected X-ray variability, with the Yohkoh Soft
X-ray Telescope, is related to its mean flux as (\cite{Katsukawa01}), ${\rm
rms}\propto\times F^{0.93\pm 0.10}$, in good agreement with the linear relationship
shown here with {\it RHESSI} data for X-ray variability including all flaring events
detected in the X-ray lightcurves. This should be compared to the case of Cygnus~X-1 in
which the X-ray variability is also mostly composed of unidentifiable flaring events,
despite that occasionally very strong flares are detectable (\cite{Gierlinski}), which
can be considered as analogy of solar flares in active regions of the Sun. Therefore
the results shown in Fig.~8 for the flux-rms linear relationship over several orders of
magnitudes of X-ray flux variations from the Sun can be regarded as a missing link
between the extremely violent flaring events from the Sun, and X-ray flux variations
with small relative amplitudes from the Sun and from accreting neutron star and black
hole binaries systems.

These results therefore demonstrate that common dynamical physical processes may
dominate the X-ray emission from the Sun, gamma-ray bursts, accreting neutron star and
black hole binaries, as well as in accreting supermassive black hole systems
(\cite{Uttley05}). Since accretion flow does not exist in the Sun, this may rule out
immediately the fluctuating accretion flow model (\cite{Lyubarskii,Kotov}) for this
type of dynamical properties occurring in neutron star and black hole binaries with
accretion disks. Because essentially all other models for X-ray emission from neutron
star and black hole systems have already been ruled out by such dynamical behaviours
(\cite{Uttley05}), new physical models for them may be required, which should have
common mechanisms to X-ray emission from the Sun. We therefore probably should return
to the solar flare-like models for accreting neutron star and black hole systems, as
well as to gamma-ray bursts.

\section{Common physical mechanisms: self-organized criticality, percolation and dynamical driving}
\begin{figure}\label{field_line}
\centerline{\includegraphics[width=5.0in]{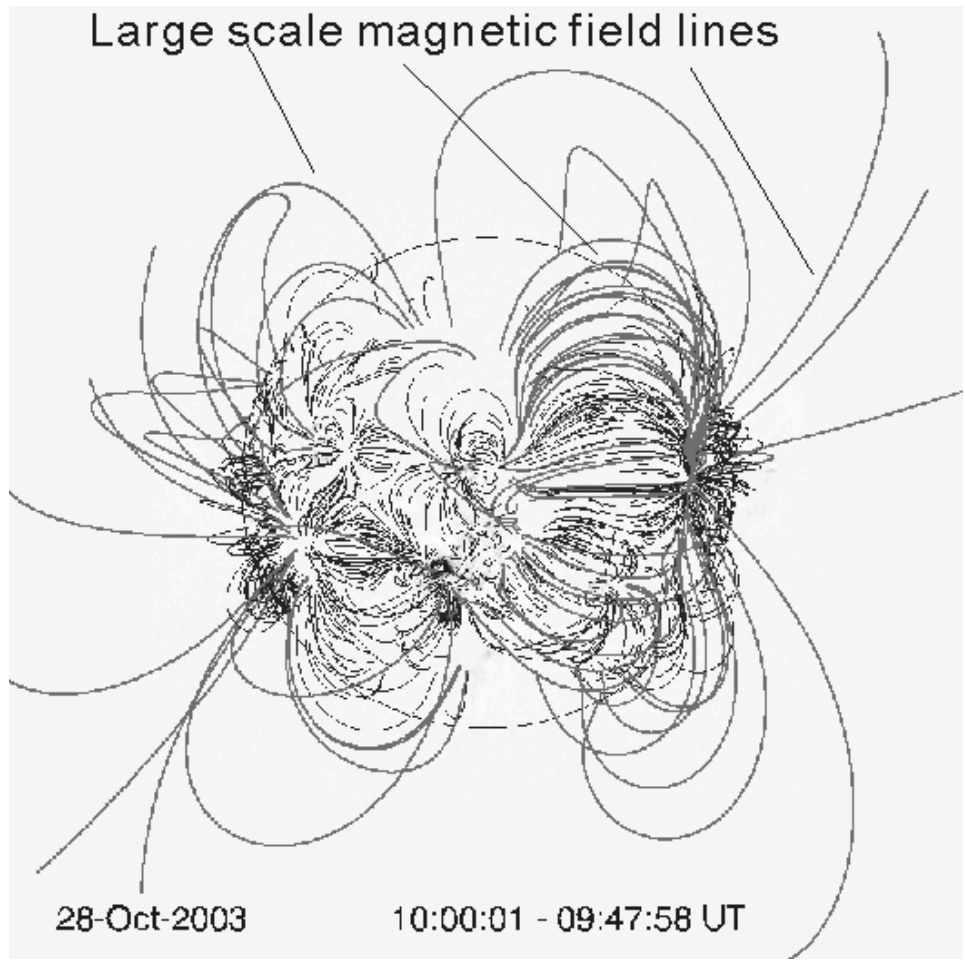} } \caption
{Solar magnetic field topology during the major solar activities on October 28th, 2003.
Many active regions are connected by complex magnetic field lines. (This figure is
adapted from \cite{wangjx}.)}
\end{figure}

\begin{figure}\label{perc}
\centerline{\includegraphics[width=4.0in]{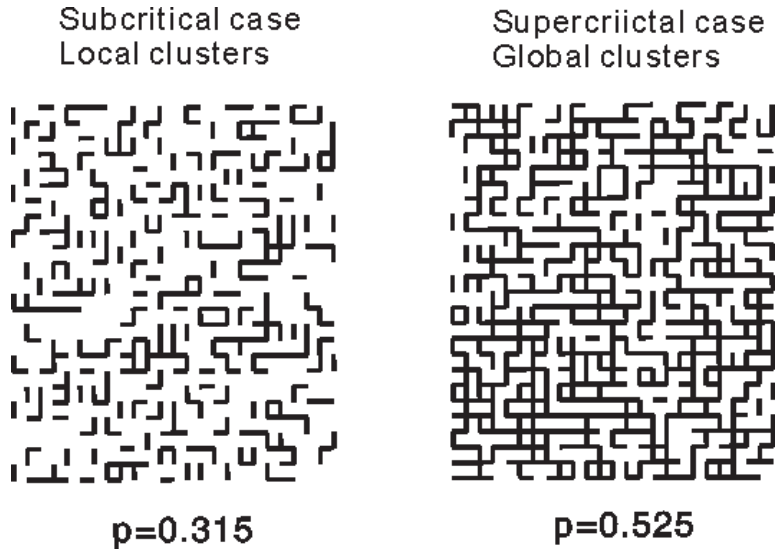} } \caption {Illustration of the
percolation model of \cite{percolation}. Two nodes are connected by an edge with
probability $p$. For $p=0.315$ (left), which is below the percolation threshold
$p_c=0.5$, the connected nodes form small and local clusters. For $p=0.525$ (right),
which is above the percolation threshold, the global and large cluster are formed.
(This figure is adapted from \cite{percolation}.)}
\end{figure}

Then what are the possible common physical mechanisms? Magnetic energy releases,
perhaps due to magnetic reconnections, should be the engines in all these systems
(\cite{Nayakshin,Matteo,Poutanen,zhang_science,liu03,wang,Liu04,Priest}). For the Sun,
 an emerging flux and reconnection model has been proposed for the
triggering of the coronal mass ejections (\cite{chen00}). In addition, some small scale
activities, such as Ellerman bombs, and type II white-light flares are possibly the
results of magnetic reconnection in the solar lower atmosphere (\cite{chen01}). The
avalanche mechanisms (\cite{Charbonneau}) driven by self-organized criticality
(\cite{Bak,Christensen}) have been widely invoked to explain previously observed
distributions of solar flare peak flux and event interval, as well as timing properties
in accreting black hole systems (\cite{Mineshige94,Takeuchi}). The multiplicative
nature of the generated events in the self-organized criticality model can produce the
observed flux-rms linear relationship. However the simple self-organized criticality
model also predicts power-law distributions of flux, which is not found in accreting
neutron star and black hole systems, and in fact also not verified satisfactorily to
the whole range of amplitudes of solar flares; the power-law flux distribution of solar
flares is the basic idea of Parker's conjecture (\cite{parker83,parker88,parker91}). As
shown in Figs.~4 \& 5, the flux distributions of individual solar flares consist of
multiple log-normal peaks, rather than a single power-law.

This mismatch between observed and predicted flux distributions may be due to
over-simplified self-organized criticality models which only describe events occurring
locally. However X-ray photons from accreting neutron star and black hole systems
cannot come from only a localized self-organized criticality spot, yet events occurred
are still inter-related as indicated by the multiplicative property. Therefore multiple
spots or a major part of the whole inner accretion disk region responsible for X-ray
emission must be inter-organized together somehow; this is further supported by the
fact that only about 30\% or less of its total X-ray flux does not follow the flux-rms
relationship in Cygnus~X-1 (see Fig.~9), i.e., only this 30\% flux is the superposition
of many randomly occurring and independent emission events. The {\it RHESSI} solar
X-ray lightcurves we analysed here are collected from the whole Sun exposed to us, and
its non-varying X-ray flux is very low, suggesting that essentially all detectable
X-ray emission in a time series over the whole Sun is produced from many inter-related
X-ray emission regions. In fact it has already been observed that many active regions
well separated and spread nearly over the whole solar surface show networked
activities; these regions are now known to be connected by large scale magnetic field
lines (\cite{wangjx}), as shown in Fig.~12.

In a simulation study of particle acceleration during solar flares, \cite{Vlahos} have
found that multi-scale magnetic fields are involved, which may be described
mathematically by the percolation model of \cite{percolation}, as illustrated in
Fig.~13. In the percolation model, two nodes are connected by an edge with a certain
probability. If the probability is below the percolation threshold, the connected nodes
form small and local clusters. In this case, most nodes will act independently, and
thus their total contributions will be additive, rather than multiplicative. On the
other hand, if the connection threshold is above the percolation threshold, global and
large clusters are formed, and thus most nodes will act inter-connectively. In this
case their total contributions will be multiplicative, because the action of each node
will influence many other nodes, even if they are well separated spatially. Such
networked activities or fluctuations can be studied by identifying the dominating
driving mechanisms, i.e., internal or external driving. If the connection probability
between two neighboring nodes is below the percolation threshold, the dynamical system
will be dominated by internal driving, or otherwise dominated by external driving.

Following \cite{external}, for a dynamical system, we can record the time dependent
activity of $N$ components, expressed by a time series $\{f_i(t)\}$, $t=1,\ldots,T$ and
$i=1,\ldots,N$. As each time series reflects the joint contribution from the system's
internal dynamics and external fluctuations, we assume that we can separate the two
contributions by writing
\begin{equation}
f_i(t)=f_i^{int}(t)+f_i^{ext}(t). \label{eq:separation}
\end{equation}
It has been shown by \cite{external} that the fluctuation or rms is linearly
proportional to $f$ if the dynamical system is externally driving, or the rms is
proportional to $\sqrt{f}$ if the dynamical system is internally driving.

Since the dominating component of the X-ray flux fluctuation or rms is proportional to
flux, for solar X-ray flares and continuous light curves, the X/gamma-ray lightcurve of
GRB~940217, accreting neutron star and black hole systems, we can infer that their
X-ray production mechanisms are all dominated by external driving dynamics. We thus
suggest that perhaps similar large scale magnetic field topology also exists in the
accretion disks around neutron stars and black holes, and also maybe in gamma-ray
bursters. We comment in passing that no current gamma-ray burst model involves
inter-connected emission nodes.  However, contrary to the widely accepted internal
shock model of prompt gamma-ray emissions for gamma-ray bursts, an alternative model,
based on electrodynamic accretion models in which the electromagnetic process turns
rotational energy into particle energy in a pulsarlike mechanism (\cite{Katz97}), has
been used to unite jets in gamma-ray bursts, accreting supermassive and stellar mass
black hole systems (\cite{Katz06}). It has also been suggested a magnetic field
dominated outflow model for GRBs is preferred, based on data from the dedicated
gamma-ray burst satellite {\it Swift} (\cite{Kumar}). These results are in qualitative
agreement with our common energy release picture for these different types of objects
at very different scales, suggesting that it may be necessary to re-examine seriously
the currently accepted standard gamma-ray burst internal shock models, by investigating
energy release processes dominated and connected by complex magnetic field topology.

Although a component of rms proportional to the square-root of total counts is
identified in all of them, especially when the total counts are small, this component
is not due to internal driving, because this component can be completely described as
Poisson counting fluctuations. This means that currently no internal driving dynamics
has been detected in any of them. It is interesting to note that the individual solar
flares show the same flux-rms linearity as for the total continuous lightcurves of all
those systems, indicating that each individual solar X-ray flare is produced from many
different inter-connected emission nodes or elements, if the percolation model is also
applicable here. It is thus reasonable to assume that regardless what triggers the
initial onset of a solar X-ray flare, the event immediately spreads out to many nodes
or elements. Imaging observations of X-ray flares have identified magnetic loops and
their foot-points as X-ray and hard X-ray production sites. Therefore it will be
important to find the relationship between the above inferred inter-connected X-ray
emission nodes or elements and observed magnetic loops and their foot-points.

\section{Concluding remarks}
In section 1, I briefly categorized different research methods in astronomy and
emphasized the importance of using the Sun as an astrophysical laboratory to study many
astrophysical phenomena across vastly different scales, but appearing to be similar to
what occurring in the Sun. I call this research discipline ``Applied Solar
Astrophysics". A beautiful example of the progress in this discipline is the use of
magnetic reconnection model by \cite {dai} to explain the late X-ray flux
re-brightening in the X-ray afterglow of a short gamma-ray burst, which is believed to
have produced a spinning young neutron star after two neutron stars merged together
(this merging event triggered the initial onset of the short gamma-ray burst).

In section 2, I briefly summarized several phenomena: (1) the atmospheric structures
around accreting black holes which are similar to solar atmosphere; (2) astrophysical
jets from many different kinds of astrophysical systems, whose production mechanisms
involve differential rotation and twisted magnetic fields similar to magnetic
reconnections in the Sun for solar flares and coronal mass ejection; (3) similar
triple-ring structure in SN1987A and Virgo cluster which are understood to be related
to rotating large scale magnetic field topology and winds, again similar to the solar
magnetic fields and solar winds. Clearly our understanding of related solar phenomena
are important for studying these similar phenomena at different scales.

In section 3, I focused on the similar non-linear dynamical properties of X-ray
variations of solar flares, a stellar mass black hole binary Cygnus~X-1 and a gamma-ray
burst GRB~940217. The remarkable consistency of the underlying non-linear dynamics
model for these three kinds of apparently very different astrophysical systems at
different scales suggests that many X-ray emission nodes or elements are somehow
inter-connected and act in response to common triggering. Again using the Sun as an
astrophysical laboratory, we suggest that complicated magnetic field topology across
many different scales in each system plays an important role in connecting many X-ray
emission nodes or elements together. It is possible that these individual nodes produce
`nano'-flares required for heating the solar corona.

However previous and current solar X-ray instruments do not have the required
capability to resolve individual X-ray flares or the quiet Sun X-ray lightcurves into
many `nano'-flares. We suggest that a future Solar X-ray Timing and Imaging ({\it
SXTI}) instrument with a large effective area and direct high resolution imaging
capability is needed to further resolve the fine timing structures of X-ray flares,
i.e., inter-connected individual `nano'-flares, to study the development of solar
flares in fine details, and overcome the Poisson counting fluctuations with a high
counting rate during the quiet Sun, in order to identify the internal driving
component, which may be made of isolated `nano'-flares.

Although it is beyond the scope of this report for proposing a conceptual design for
the {\it SXTI} telescope, here we outline only the basic requirements: (1) an effective
area of $>$100 cm$^2$ at a few keV to be compared with the 30 cm$^2$ effective area at
10 keV for {\it RHESSI}, however {\it RHESSI} relies on rotation modulation imaging and
thus cannot offer simultaneous imaging and high timing resolution capability; (2)
direct imaging angular resolution of $<$ 1 arcsec, to be compared with HINODE
(Solar-B)'s X-ray
telescope\footnote{http://solar-b.nao.ac.jp/xrt\_e/fact\_sheet\_e.shtml} which has 2
arcsec direct imaging resolution, but with only 1 cm$^2$ effective area at 0.523 keV
and 2 second timing resolution; and (3) timing resolution $<$ 10 ms which requires
advanced X-ray CCDs, such as that in current developments (e.g., \cite{zhang_chen}).

For comparison the currently operating
Chandra X-ray telescope\footnote{http://chandra.harvard.edu/about/science\_instruments.html}
 has a peak effective area of about 800 cm$^2$ at around 2 keV,
with a direct imaging resolution of 0.5 arcsec and each X-ray CCD's full-frame readout
time of around 3.3 seconds. Although telescopes pointed at the Sun involve
significantly more complex technical difficulties, it is possible that a solar X-ray
observatory with the above proposed {\it SXTI} will be technically feasible within the
near future. With such an instrument, our understanding of solar X-ray production
mechanism, and subsequently many currently open issues on solar magnetic fields,
particle acceleration and coronal heating, etc, will be advanced significantly. Because
of many astrophysical phenomena at different scales should have similar underlying
physical processes to what occurring in the Sun, we should also make significant
progresses towards understanding many important astrophysical problems. Therefore
``Applied Solar Astrophysics" has a bright future.

\begin{acknowledgments}
I shall thank Cheng Fang of Nanjing University for nominating me, the IAU executive
committee for inviting me as one of the four discourse speakers at the 26th IAU GA in
Prague, Czech Republic, August 2006,  and Ding-Qiang Su of Nanjing University, who was
then the President of the Chinese Astronomical Society, for encouragements and advises
during my preparation of this invited discourse. I also appreciate discussions with Bob
Lin of UC Berkeley and Don Melrose of University of Sydney during the 26th IAU GA on
{\it RHESSI} data analysis and solar particle accelerations, respectively. The many
kind compliments and stimulating discussions from the audience in the discourse are the
best rewards for my discourse, which has indeed taken me a lot of time and efforts to
prepare for.

I also thank Shui Wang of University of Science and Technology of China for discussions
on magnetic reconnections, Jun-Han You of Shanghai Jiao-Tong University for discussions
on radiation mechanisms, Wei-Qun Gan of Purple Mountain Observatory for advises on {\it
RHESSI} data analysis and conversations on solar flares, Jing-Xiu Wang of National
Astronomical Observatories of China and Louise Harra of Mullard Space Science
Laboratory of University College London for exchanges on solar physics, Jian-Min Wang
of Institute of High Energy Physics for many discussions on accretion physics, and many
of my previous collaborators and formal students for the privilege and fun of working
with them and for their important contributions to our works discussed and/or
referenced in this report. My student Shi-Cao Tang of Tsinghua University is greatly
acknowledged for assisting me in most of the data analysis work shown in Figs.~3-11 of
this report. Finally Cheng Fang, Zi-Gao Dai \& Peng-Fei Chen of Nanjing University, Tan
Lu of Purple Mountain Observatory, Jun Lin of Yunnan Observatory, Jing-Xiu Wang of
National Astronomical Observatories of China, Phil Uttley of Southampton University and
Simon Vaughan of Leicester University are thanked for comments and suggestions on the
manuscript of this report.

This work is supported in part by the Ministry of
Education of China, Directional Research Project of the Chinese Academy of Sciences and
by the National Natural Science Foundation of China under project no.
10521001, 10327301, 10233010 and 10233030.

\end{acknowledgments}


\end{document}